\def\plushc{\ +\ h.c.}
\def\diag{{\rm diag}}
\def\sgn{{\rm sgn}}
\newcommand{\nn}{\nonumber}

\def\be{\begin{equation}}
\def\ee{\end{equation}}
\def\bea{\begin{eqnarray}}
\def\eea{\end{eqnarray}}
\def\Ref#1{(\ref{#1})}
\def\cropen#1{\crcr\noalign{\vskip #1}}
\def\crr{\cropen{1\jot }}
\newcommand{\rc}{r}
\newcommand{\keff}{\kappa_{{\rm eff}}}
\newcommand{\I}{{\rm i}}
\newcommand{\hfive}{{\hat 5}}

\documentstyle[epsf, 12pt]{article}
\begin{document}

\thispagestyle{empty}
\begin{flushright}
CIT-USC/00-015\\
\end{flushright}

\bigskip\bigskip\begin{center} {\bf
\LARGE{\bf Supersymmetric Randall-Sundrum Scenario} }
\end{center} \vskip 1.0truecm

\centerline{\bf Richard Altendorfer, Jonathan Bagger}

\vskip5mm
\centerline{\it Department of Physics and Astronomy}
\centerline{\it The Johns Hopkins University}
\centerline{\it 3400 North Charles Street}
\centerline{\it Baltimore, MD 21218}
\vskip5mm
\centerline{\bf Dennis Nemeschansky}

\vskip5mm
\centerline{\it CIT-USC Center for Theoretical Physics and}
\centerline{\it Department of Physics and Astronomy}
\centerline{\it University of Southern California}
\centerline{\it Los Angeles, CA 90089}
\vskip5mm

\bigskip \nopagebreak \begin{abstract}
\noindent We present the supersymmetric version of the minimal
Randall-Sundrum model with two opposite tension branes.
\end{abstract}

\newpage\setcounter{page}1

\section{Introduction}

The two-brane
Randall-Sundrum scenario \cite{RS} provides an appealing
way to generate the electroweak gauge hierarchy as a
consequence of spacetime geometry.  The basic idea is to
start with five dimensional anti de Sitter (AdS) space,
take the region between two slices parallel to the AdS
horizon, and add a 3-brane along each slice.  By tuning
the brane tensions, the resulting configuration can be
made stable against gravitational collapse.

In this model, the ratio of the weak to the Planck
scale is determined by the distance between the two branes.
The distance is fixed by the expectation value of a modulus
field, called the radion.  The usual hierarchy problem now
appears in a new guise:  What fixes the radion vev, and
what protects the vev against large radiative corrections?

In a recent paper, Goldberger and Wise \cite{GW} proposed
a way to stabilize the radion using five dimensional bulk
matter.  Supersymmetry provides another possibility.  In
this paper we will take some first steps in that direction,
and show how to supersymmetrize the minimal Randall-Sundrum
scenario.

In what follows we will use coordinates in which the five
dimensional background metric takes the following form,
\be
ds^2\ =\ e^{- 2\sigma(\phi)}\, \eta_{mn}\, dx^m dx^n
\ +\ \rc^2 \, d\phi^2\ .
\label{AdSmetric}
\ee
The coordinate $x^5 = \rc \phi$ parametrizes an orbifold
$S^1/Z_2$, where the circle $S^1$ has radius $\rc$ and
the orbifold identification is $\phi \leftrightarrow -\phi$.
For fixed $\phi$, the coordinates $x^m$ ($m = 0,1,2,3$)
span flat Minkowski space, with metric $\eta_{mn} =
\diag(-1,1,1,1)$.  We choose to work on the orbifold
covering space, so we take $-\pi < \phi \le \pi$.

For the gravitational part of our action, we follow
Randall and Sundrum and take the action to be the sum
of bulk plus brane pieces,
\be
S\ =\ S_{\rm bulk}\ +\ S_{\rm brane}\ .
\label{totalS}
\ee
The bulk action is that of pure five dimensional AdS gravity,
while $S_{\rm brane}$ arises from the presence of two
opposite tension branes.

The gravitational bulk action is given by
\be
S_{\rm bulk}\ =\ {\Lambda \over \kappa^2}\,\int d^5x \,e\,
\left[\, - {1 \over 2}\,R\ +\ 6\, \Lambda^2 \, \right]\ ,
\label{bulkS}
\ee
where $\kappa$ is related to the four dimensional Planck constant,
$e = \det e_M{}^A$, and $e_M{}^A$ is the five dimensional
f\"unfbein.\footnote{We adopt the convention that capital
letters run over the set $\{0,1,2,3,5\}$ and lower-case
letters run from 0 to 3.  Tangent space indices are taken
from the beginning of the alphabet; coordinate indices
are from the middle.  We follow the conventions of
\cite{WB}.}  In this expression, $\Lambda$ is
the bulk cosmological constant and $R$ is the five
dimensional Ricci scalar,
\be
R\ =\ e_A{}^M e_B{}^N \, R_{MN}{}^{AB}\ .
\ee
The Riemann curvature $R_{MN}{}^{AB}$ is built from the
spin connection according to the following conventions,
\be
R_{MN}{}^{AB}\ =\ \partial_M \omega_N{}^{AB}\ -\ \partial_N
\omega_M{}^{AB}\ -\ \omega_M{}^{AC}\omega_{NC}{}^{B}
\ +\ \omega_N{}^{AC}\omega_{MC}{}^{B}\ .
\ee

The brane action serves as a source for the bulk gravitational
fields.  It arises from the 3-branes located at the orbifold
points $\phi = 0, \pi$.  For the case at hand, the brane action
is simply
\be
S_{\rm brane}\ =\ -6\ {\Lambda^2 \over \rc \kappa^2}
\ \int d^5x \,\hat{e}\,
\left[\,\delta(\phi)\, -\, \delta(\phi-\pi) \,\right]\ ,
\label{braneS}
\ee
where $\hat{e} = \det e_m{}^a$, and the $e_m{}^a$ are the
components of the five dimensional f\"unfbein, restricted to
the appropriate brane.

{}From this action it is not hard to show that the metric
\Ref{AdSmetric}, with
\be
\sigma(\phi)\ =\ \rc\,\Lambda\,|\phi|\ ,
\ee
is a solution to the five dimensional Einstein equations,
\be
R_{MN}\ -\ {1\over2} g_{MN} R\ =\ -6\,  g_{MN}\,\Lambda^2\ +
\ 6 \,g_{mn}\,\delta^m_M
\delta^n_N  \left( {\Lambda\over \rc}\right)
\left( {\hat e \over e}\right)
[\,\delta(\phi)\, -\, \delta(\phi-\pi)\,]\ .
\ee
Away from the branes, the bulk metric is just that of
five dimensional AdS space, with cosmological constant
$\Lambda$.  On the branes, the four dimensional metric
is flat.  As shown in \cite{RS}, the effective theory of
the gravitational zero modes is just ordinary four
dimensional Einstein gravity, with a vanishing cosmological
constant.  The effective four dimensional squared Planck mass
is $\keff^{-2} = \kappa^{-2}(1 - e^{-2 \pi \rc\Lambda})$.

\section{Supersymmetric Bulk}

In what follows we will supersymmetrize the action
\Ref{totalS}.  We start with the bulk action \Ref{bulkS}.
Its supersymmetric extension can be found from the five
dimensional supersymmetric AdS action \cite{AdS},
\bea
S_{\rm bulk} & = & \Lambda\, \int d^5x
\,e\,\bigg[\,- {1 \over 2 \kappa^2} R
\ + \I \epsilon^{MNOPQ}
\overline\Psi_{M}\Sigma_{NO} D_P \Psi_Q
\ -\ {1 \over 4} F_{MN} F^{MN} \nn \\
& & - \ 3\, \Lambda \,\bar\Psi_M\Sigma^{MN}\Psi_N
\ +\ 6 \,{\Lambda^2\over\kappa^2}
\ -\ \I \kappa {\sqrt{3\over2}}{1\over2}
F_{MN}\bar\Psi^M\Psi^N \nn\\
& &-\ \kappa {1\over 6\sqrt{6}}
\epsilon^{MNOPQ}F_{MN}F_{OP}B_Q
\ +\ \I \kappa \sqrt{3\over2}{1\over 4}
\epsilon^{MNOPQ}F_{MN}
\bar\Psi_O \Gamma_P\Psi_Q \nn\\
& & -\ \kappa\, \Lambda {\sqrt{3\over 2}}
\epsilon^{MNOPQ}
\bar\Psi_M\Sigma_{NO}\Psi_P B_Q
\ +\ \hbox{ four-Fermi\ terms } \bigg]\ ,
\eea
where the $\epsilon$ tensor is defined to have tangent-space
indices, and $\epsilon^{01235} = 1$.
This action contains the physical fields associated with
the supergravity multiplet in five dimensions:  the f\"unfbein
$e_M{}^A$, the gravitino $\Psi_M$, and a vector field $B_M$.
The covariant derivative $D_M \Psi_N = \partial_M \Psi_N
+ {1\over2} \Sigma^{AB} \omega_{MAB} \Psi_N$ and the matrix
$\Sigma^{AB} = {1\over4}(\Gamma^A \Gamma^B - \Gamma^B
\Gamma^A)$.

This action is invariant under the following supersymmetry
transformations,
\bea
\delta e_M{}^A &=& \I\kappa\, (\bar\eta\Gamma^A\Psi_M\ -
\ \bar\psi_M\Gamma^A \eta) \nn\\
\delta B_M &=& -\I\sqrt{3\over 2}\,(\bar\eta\psi_M\ -
\ \bar\psi_M\eta) \nn\\
\delta \psi_M &=& {2\over\kappa}D_M\eta
\ +\ {\I}\,{\Lambda\over\kappa}\Gamma_M\eta
\ -\ \I\sqrt{6}\,\Lambda B_M \eta \nn\\
& & -\ \sqrt{2\over 3}\,(\Gamma^N F_{NM}\ -\ {1\over4}
\epsilon_{MNOPQ}F^{NO}\Sigma^{PQ})\eta \nn\\[2mm]
& & +\ \hbox{ three-Fermi\ terms. }
\label{susytrans}
\eea
Since we work in the orbifold covering space, the spacetime
manifold has no boundary, and we can freely integrate by
parts.  We use the 1.5 order formalism, so the spin
connection obeys its own equation of motion and does not
need to be varied.

For the case at hand, we must define the action of the
orbifold symmetry on the AdS fields.  We start by writing
the five dimensional spinors in a four dimensional language,
where
\be
\Psi_M \ \rightarrow\ \pmatrix{ \psi^1_{M \alpha} \crr
\bar\psi^{\dot\alpha}_{2 M}}
\ee
and
\be
\Gamma^a \ \rightarrow\ \pmatrix{ 0 &
\sigma^a_{\alpha\dot\alpha} \crr
\bar\sigma^{a\dot\alpha\alpha} & 0} \qquad \qquad
\Gamma^5\ \rightarrow\ \pmatrix{ -\I & 0 \crr 0 & \I}\ .
\ee
The fields $\psi^i_M$ (for $i = 1,2$) are two-component
Weyl spinors, in the notation of \cite{WB}.  We
then define $\psi^\pm_M = {1\over \sqrt 2}(\psi^1_M \pm
\psi^2_M)$, and likewise for $\eta^\pm$.

In terms of these fields, the bulk supersymmetry transformations
can be written in the following form,
\bea
\delta e_M{}^a &=& \I\kappa\, (\eta^+\sigma^a\bar\psi^+_{M}
\, +\, \eta^-\sigma^a\bar\psi^-_{M})
\plushc \nn\\[2mm]
\delta e_M{}^\hfive &=& \kappa\, (\eta^+\psi^-_{M}
\, -\, \eta^-\psi^+_{M})
\plushc \nn\\
\delta B_M &=& -\I\sqrt{3\over 2}\,
(\eta^+\psi^-_{M}
\, -\, \eta^-\psi^+_{M})  \plushc  \nn\\
\delta \psi^\pm_m &=& {2\over\kappa}\,D_m\eta^\pm
\ \mp\ {\I\over\kappa}\, \omega_{ma\hfive} \,\sigma^a
\bar\eta^\mp \ \pm\ {\I}\,{\Lambda\over\kappa}
\,e_m{}^a\, \sigma_a \bar\eta^\pm
\ +\ \,e_{m\hfive}\,{\Lambda\over\kappa}\,\eta^\mp \nn\\
&& -\ \I\,\sqrt{6}\,\Lambda\, B_m\, \eta^\mp
\ -\ \sqrt{2\over 3}\,\bigg(\mp \,e_a{}^N\,F_{Nm}\,
\sigma^a\bar\eta^\mp\ -\ \I\, e_\hfive{}^N\, F_{Nm}\, \eta^\pm \nn\\
&& -\ {1\over 4}\, \epsilon_{ABCde}\,
e_m{}^A e^{BN} e^{CO} \,F_{NO} \,\sigma^{de} \eta^\pm
\ \pm\ {\I\over 4}\, \epsilon_{abcd}\,
e_m{}^a e^{bN} e^{cO} \,F_{NO} \,\sigma^{d} \bar\eta^\mp \bigg)
\nn\\[2mm]
\delta \psi_5^\pm &=& {2\over\kappa}\,D_5\eta^\pm
\ \mp\ {\I\over\kappa}\, \omega_{5a\hfive}
\,\sigma^a \bar\eta^\mp
\ +\ e_{5\hfive}\,{\Lambda\over\kappa}\, \eta^\mp
\ \pm\ \I\,e_5{}^a\,{\Lambda\over\kappa}\,
\sigma_a \bar\eta^\pm \nn\\
&& -\ \I\sqrt{6}\,\Lambda\, B_5 \,\eta^\mp
\ +\ \sqrt{2\over 3}\,\bigg(\mp \,e_a{}^n\,F_{5n}\,
\sigma^a\bar\eta^\mp\ -\ \I\, e_\hfive{}^n\, F_{5n}\, \eta^\pm \nn\\
&& +\ {1\over 4}\, \epsilon_{ABCde}\,
e_5{}^A e^{BN} e^{CO} \,F_{NO} \,\sigma^{de} \eta^\pm
\ \pm\ {\I\over 4}\, \epsilon_{abcd}\,
e_5{}^a e^{bN} e^{cO} \,F_{NO} \,\sigma^{d} \bar\eta^\mp \bigg)\ .
\label{susytrans2}
\eea
In these expressions, all fields depend on the five dimensional
coordinates.  The symbol $\hfive$ denotes the fifth tangent space
index, and all covariant derivatives contain the spin connection
$\omega_{Mab}$.  Here and hereafter, we ignore all three- and
four-Fermi terms.

{}From these transformations it is not hard to find a consistent
set of $Z_2$ parity assignments under the orbifold transformation
$\phi \rightarrow -\phi$.  The assignments must leave the action
and transformation laws invariant under the $Z_2$ symmetry.  We
assign even parity to
$$
e_m{}^a, \quad e_{5\hfive}, \quad B_5, \quad \psi^+_{m}, \quad
\psi^-_{5}, \quad \eta^+
$$
and odd parity to
$$
e_5{}^a, \quad e_m{}^\hfive, \quad B_m, \quad \psi^-_{m}, \quad
\psi^+_{5}, \quad \eta^- \ .
$$

The bulk supergravity action is invariant under $N=1$ supersymmetry
in five dimensions.  The branes break all but one
one four dimensional supersymmetry.
To find its form, we shall study the supersymmetry transformations
in the orbifold background, where $e_{5\hfive} = 1$, $e_m{}^a =
e^{-\sigma(\phi)}\,\delta^a_m$, and all other fields equal zero.  This
configuration satisfies the gravitational equations of motion when
$\sigma(\phi) = \rc \Lambda |\phi |$.  Note that this background
is consistent with the orbifold symmetry.

In the orbifold background, the supersymmetry variations of the
bosonic fields are obviously zero.  The variations of the fermions
are a little trickier.  In this background, the spin connection
evaluates to
\be
\omega_{mAM}\Sigma^{AM}\ =\ \sgn(\phi)\,\Lambda\,
\Gamma_m\Gamma^\hfive\ ,
\ee
with all other components zero.  The supersymmetry variations
of the fermions reduce to the following form,
\bea
\delta \psi^\pm_m &=&  {2\over\kappa}\,\partial_m\eta^\pm
\ \mp\ {\I}\, \sgn(\phi)\, {\Lambda\over\kappa}
\, \sigma_m \bar\eta^\mp
\ \pm\ {\I}\, {\Lambda\over\kappa}
\, \sigma_m \bar\eta^\pm\nn\\
\delta \psi^\pm_5 &=&  {2\over\kappa}\,\partial_5\eta^\pm
\ +\  {\Lambda\over\kappa}\,\eta^\mp\ .
\eea

The unbroken supersymmetries are found by setting
these variations to zero.  The resulting Killing equations
can then be solved for the Killing spinors $\eta^\pm$.
The solution that reduces to a flat-space supersymmetry
in four dimensions is simply
\be
\eta^+\ =\ {1\over\sqrt2}\, e^{- \sigma(\phi)/2}\,\eta(x)\ ,
\qquad\qquad
\eta^-\ =\ {1\over\sqrt2}\, e^{- \sigma(\phi)/2}\,\sgn(\phi)
\eta(x)\ ,
\label{etazeromode}
\ee
where $\sgn(\phi)$ is the step function,\footnote{The
distribution $\sgn(\phi)$ obeys the following
properties:
$$
\int^\epsilon_{-\epsilon} d\phi\ \sgn(\phi)
\ =\  0 \ ,\qquad
\int^\epsilon_{-\epsilon} d\phi\ \sgn^2(\phi)
\ =\  2 \epsilon \ ,
$$
when integrated against smooth functions, and
$$
\int^\epsilon_{-\epsilon}
d\phi\ \sgn(\phi)\, \delta(\phi) \ =\  0 \ , \qquad
\int^\epsilon_{-\epsilon} d\phi\ \sgn^2(\phi)
\,\delta(\phi) \ =\ {1\over3} \ ,
$$
when integrated against $\delta(\phi)$. The last
relation ensures that
$$
\int^\epsilon_{-\epsilon} d\phi
\ {d\over d\phi}\,\sgn^3(\phi) \ =\ 2\ .
$$
We thank Jan Conrad for a discussion on this point.}
which evaluates to $(-1,0,1)$, depending on the sign of $\phi$.
In this expression,
the spinor $\eta$ contains four Grassmann components
and is a function of $x^0,...,x^3$, but not $x^5$.  We
shall see that it describes the one unbroken supersymmetry
of the Randall-Sundrum scenario.

It is not hard to check that the spinors \Ref{etazeromode}
are a solution to the Killing equations, for constant $\eta$,
except for delta-function singularities at the orbifold points
$\phi = 0, \pi$.  These singularities are very important. They
motivate us to change the $\psi^-_5$ supersymmetry transformation
so that the spinors \Ref{etazeromode} are Killing spinors
everywhere.  We take
\bea
\delta \psi_5^- &=& {2\over\kappa}\,D_5\eta^-
\ +\ {\I\over\kappa}\, \omega_{5a\hfive}
\,\sigma^a \bar\eta^+
\ +\ e_{5\hfive}\,{\Lambda\over\kappa}\, \eta^+
\ -\ \I\,e_5{}^a\,{\Lambda\over\kappa}\,
\sigma_a \bar\eta^- \nn\\
&& -\ \I\sqrt{6}\,\Lambda\, B_5 \,\eta^+
\ +\ \sqrt{2\over 3}\,\bigg( \,e_a{}^n\,F_{5n}\,
\sigma^a\bar\eta^+\ -\ \I\, e_\hfive{}^n\, F_{5n}
\, \eta^- \nn\\
&& +\ {1\over 4}\, \epsilon_{ABCde}\,
e_5{}^A e^{BN} e^{CO} \,F_{NO} \,\sigma^{de} \eta^-
\ -\ {\I\over 4}\, \epsilon_{abcd}\,
e_5{}^a e^{bN} e^{cO} \,F_{NO} \,\sigma^{d} \bar\eta^+
\bigg) \nn\\
&& -\ {4\over \rc\kappa}
\, [\,\delta(\phi)\, -\, \delta(\phi-\pi)\,]\,
\eta^+ \ .
\label{newtrans}
\eea
In the orbifold background, this reduces to
\be
\delta \psi^-_5 \ =\   {2\over\kappa}\,\partial_5\eta^-
\ +\  {\Lambda\over\kappa}\,\eta^+
\ -\ {4\over \rc\kappa}\,[\,\delta(\phi)\, -\,
 \delta(\phi-\pi)\,]\,\eta^+\ .
\ee
The spinors \Ref{etazeromode} satisfy the modified Killing
equations, for constant $\eta$, even at the orbifold points
$\phi = 0, \pi$.  Furthermore, the supersymmetry
transformations still close into the $N=1$ supersymmetry
algebra.

\section{Supersymmetric Brane}

In the previous section, we changed the gravitino
supersymmetry transformations so that the Killing
spinors satisfy the Killing equations at every point
in $\phi$.  Because of this, the bulk action is no
longer invariant under the supersymmetry transformations
\Ref{newtrans}.  In this section we will find a brane
action whose variation precisely cancels that of the
bulk.

We first compute the variation of the
bulk action.  Comparing \Ref{susytrans2} with
\Ref{newtrans}, we see that the bulk variation
vanishes except on the branes.  Therefore, to
compute the variation, we need to project the bulk
fields onto the branes.  For even fields, this is easy:
The brane fields are just the bulk fields evaluated at
the appropriate value of $\phi$.   For odd fields,
the situation is more subtle:  The brane fields must
obey jump conditions across the delta function
singularities and these conditions are determined
by the brane action.

In what follows we will present the brane action
and verify that it restores the supersymmetry of
the bulk-plus-brane system.  We assert that the
brane action is simply
\be
S_{\rm brane} \ =\ {\Lambda \over \rc \kappa^2 }
\, \int d^5x \,\hat{e}
\ (-3 \Lambda
\ +\ 2\,\kappa^2\, \psi^+_m \sigma^{mn} \psi^+_n)
\ \left[\,\delta(\phi) - \delta(\phi-\pi) \,\right]
\plushc
\label{braneS2}
\ee
where the fields $e_m{}^a$ and $\psi_m^+$ are projections
of the corresponding five dimensional fields.

Given this brane action, it is easy to compute
the jump conditions.  From the equations
of motion for $e_m{}^a$ and $\psi^+_m$, we find
\be
[\omega_{ma5}]\ =\ \pm\,2 \Lambda\,e_{ma}
\ , \qquad
[\psi^-_m]\ =\ \pm\,2 \psi^+_m\ ,
\label{jump}
\ee
where the square brackets denote the discontinuity across
the singularity, and the $\pm$ applies to the brane at
$\phi = 0$ and $\pi$, respectively.  A consistent solution
is given by
\be
\omega_{ma5}\ =\ \sgn(\phi)\,\Lambda\,e_{ma}
\ ,\qquad \psi^-_m\ =\ \sgn(\phi)\,\psi^+_m\ ,
\label{jumpsol}
\ee
in the neighborhood of the branes.  All other odd fields
vanish on the branes.

Now that we have the solutions to the jump conditions,
we are free to compute the variation of the bulk action.
A small calculation gives
\bea
\delta S_{\rm bulk} &=& {\Lambda\over \rc \kappa }\,
\int d^5x \,e\,e^{\hfive 5} \bigg[\,\bigg(
\,8\,\eta^+\sigma^{mn}D_m \psi^+_n\, -\,
\I\,\kappa\,
\sqrt{6}\,F^{\hfive m}\,\eta^+\psi^+_m \nn\\[2mm]
&& +\ 6\I\,\Lambda\,(1-\sgn^2(\phi))\,
\eta^+\sigma^m\bar\psi^+_{m}\,\bigg)
\ \left[\,\delta(\phi) - \delta(\phi-\pi) \,\right]
\bigg]\ \plushc
\label{deltabulk}
\eea
where $\eta^+$ is the spinor \Ref{etazeromode}.  In what
follows we will show that the variation of the brane action
precisely cancels this term.

The supersymmetry variation of the brane action is not
hard to find.  The supersymmetry transformations are
those of the bulk fields, as projected on the branes,
subject to the jump conditions \Ref{jump}.  From
\Ref{susytrans2} and \Ref{jumpsol}, we compute
\bea
\delta e_m{}^a &=& \I\kappa\,(1+\sgn^2(\phi))\,
\eta^+\sigma^a\bar \psi^+_{m} \plushc \nn\\[1mm]
\delta \psi^+_m &=& {2\over\kappa}\,D_m\eta^+
+\ {\I}\,{\Lambda\over\kappa}\,(1-\sgn^2(\phi))
\,e_m{}^a\, \sigma_a \bar\eta^+\nn\\[2mm]
&& +\ \I\,\sqrt{2\over 3}\, F_{\hfive m}\, \eta^+
\ +\ \I\, \sqrt{2\over 3}\,
F^{\hfive n}\, \sigma_{mn} \eta^+\ .
\label{branetrans}
\eea
As above, $\eta^+$ is given by \Ref{etazeromode}.
In all fields, the coordinate $\phi$ is evaluated at
$\phi = 0$ or $\pi$, depending on the location of
the brane.  Substituting \Ref{branetrans} into
\Ref{braneS2}, we find
\bea
\delta S_{\rm brane} &=& -{\Lambda\over \rc \kappa }\,
\int d^5x \,\hat{e}\,\bigg[\,\bigg(
\,8\,\eta^+\sigma^{mn}D_m \psi^+_n\, -\,
\I\,\kappa\,
\sqrt{6}\,F^{\hfive m}\,\eta^+\psi^+_m \nn\\[2mm]
&& +\ 12\I\,\Lambda\,\sgn^2(\phi)\,
\eta^+\sigma^m\bar\psi^+_{m}\,\bigg)
\ \left[\,\delta(\phi) - \delta(\phi-\pi) \,\right]
\bigg]\ \plushc
\label{deltabrane}
\eea
The variation of the brane action, \Ref{deltabrane},
cancels the variation of the bulk action, \Ref{deltabulk},
because $e =e_{5\hfive}\,\hat e $ and $\sgn^2(\phi) =
1/3$ when integrated against a delta function.
This proves that the full bulk-plus-brane Randall-Sundrum
action is invariant under the four dimensional
supersymmetry parametrized by the Killing spinor $\eta$ in
Eq.~\Ref{etazeromode}.

\section{Minimal Effective Action}

We will now derive the effective four dimensional action
for the supergravity zero modes.  We will see that it
is nothing but the usual on-shell four dimensional
flat-space supergravity action.

The zero modes of the four dimensional theory must satisfy
the massless equations of motion in four dimensions.  For
the vierbein, the zero mode was given by Randall and Sundrum
\cite{RS}:
\be
e_M{}^A\ =\ \pmatrix{ \quad 1\quad & 0 \cr
0 & e^{- \sigma(\phi)}  \bar{e}_m{}^a(x)  }\ ,
\ee
where $\sigma(\phi) = \rc \Lambda |\phi|$ and the vierbein
$\bar{e}_m{}^a$ is a function of $x^0,...,x^3$, but not
$x^5$.  The five dimensional Einstein equations, with brane
sources, reduce to the usual four dimensional source-free
Einstein equations for the vierbein $\bar{e}_m{}^a$.

The gravitino zero modes can be found in a similar way.  One
starts with the five dimensional gravitino equations of
motion,
\bea
\partial_5\psi^+_m\ +\ {3\over 2}\,\Lambda\,\psi^-_m
\ -\ \sgn(\phi) \Lambda\,\psi^+_m
 &=& 0 \nn\\
\partial_5\psi^-_m\ +\ {3\over 2}\,\Lambda\,\psi^+_m
\ -\ \sgn(\phi) \, \Lambda\psi^-_m
&=& {2\over \rc}\, \left[\,\delta(\phi) - \delta(\phi-\pi)
\,\right]\, \psi^+_m \ ,
\label{gravitino}
\eea
and assumes the following ansatz,
\be
\psi^+_m\ =\ {1\over \sqrt 2}\,
\left({\keff\over\kappa}\right)\,e^{-\sigma(\phi)/2}\,
\psi_m(x)\ ,\qquad
\psi^-_m\ =\ {1\over \sqrt 2}\,
\left({\keff\over\kappa}\right)\,e^{-\sigma(\phi)/2}\,
\sgn(\phi)\,\psi_m(x)\ .
\label{gravitinoansatz}
\ee
Substituting \Ref{gravitinoansatz} into \Ref{gravitino}, one
recovers the usual four dimensional equations of motion for the
gravitino field $\psi_m$.

In what follows, we will derive the effective four dimensional
action for the supergravity zero mode fields.  We start by
setting all other fields to zero.  This truncation is consistent
with the supersymmetry transformations \Ref{susytrans}.  We then
substitute the zero-mode expressions into the supersymmetric
bulk-plus-brane action and integrate over the coordinate
$x^5$.  We use the fact that
\be
R\ =\ e^{2\sigma} \bar{R}\ +\ 20 \Lambda^2\ -\
16 {\Lambda\over \rc} \,[\delta(\phi) - \delta(\phi-\pi)]
\ee
and
\be
\omega_{mAB}\Sigma^{AB}\ =\ \sgn(\phi)\,\Lambda
\Gamma_m \Gamma^\hfive\ +\ \bar{\omega}_{mab} \sigma^{ab}
\ee
to find
\be
S_{\rm eff} \ =\ \int d^4x
\,\bar{e}\,\bigg[\,- {1 \over 2 \keff^2} \bar{R}
\ + \ \epsilon^{mnpq}
\,\overline\psi_{m}\bar{\sigma}_{n} D_p \psi_q \,\bigg]\ ,
\ee
up to four-Fermi terms.
This is nothing but the on-shell action for flat-space $N=1$
supergravity in four dimensions.

The supersymmetry transformation laws can be found in a
similar way.
We start with the supersymmetry transformation parameters
$\eta^+$ and $\eta^-$ as above, in \Ref{etazeromode}.  We then
substitute the zero mode expressions into the supersymmetry
transformations \Ref{susytrans}.  All $x^5$ dependent terms
cancel, leaving
\bea
\delta e_m{}^a &=& \I\keff\, \eta \sigma^a\bar
\psi_{m} \plushc \nn\\[1mm]
\delta \psi_m &=& {2\over\keff}\,D_m\eta\ .
\eea
These are nothing but the transformations of $N=1$ supergravity
in four dimensions (up to three-Fermi terms), with an effective
four dimensional squared Planck
mass, $\keff^{-2} = \kappa^{-2}(1 - e^{-2 \pi \rc\Lambda})$.

\section{Summary and Outlook}

In this paper we supersymmetrized the minimal Randall-Sundrum
scenario.  We found the supersymmetric bulk-plus-brane action
in five dimensions, as well as the corresponding supersymmetry
transformations.  We solved for the Killing spinor that describes
the unbroken $N=1$ supersymmetry of the four dimensional effective
theory.  We derived the supergravitational zero modes, and showed
that the low energy effective theory reduces to ordinary $N=1$
supergravity in four dimensions.

This work represents a first step towards a deeper understanding
of supersymmetry in the context of warped compactifications.
To study stability, one would like, of course, to include the
radion multiplet, which reduces to $N=1$ matter in four
dimensions.  For phenomenology, one would also like to add
supersymmetric matter on the branes and in the bulk.  Work
along all these lines is in progress.

We are pleased to acknowledge helpful conversations with Jan
Conrad, Dan Freedman, Erich Poppitz, Raman Sundrum and Max
Zucker.   This work was supported by the National Science
Foundation, grant NSF-PHY-9404057, and the Department of
Energy, contract DE-FG03-84ER40168.

\vspace{0.25in}
{\it Note added:} On the same day this paper was submitted
to the archive, a similar paper was posted by Gherghetta
and Pomarol \cite{GP}.  This paper used an $x^5$-dependent
bulk gravitino mass to supersymmetrize the two-brane
Randall-Sundrum scenario.  The resulting construction
can be interpreted as a truncation of a more fundamental
theory with matter in the bulk.  We did not take this
approach because our goal was to supersymmetrize the purely
gravitational case.  For more on the difficulties of constructing
brane-like solutions in matter-coupled five dimensional
supergravity, see \cite{KL} and \cite{BC}.

\end{document}